\documentclass[%
 reprint,%
 amssymb, amsmath,%
 aip,cha,%
]{revtex4-1}
\pdfoutput=1
\usepackage{float}
\usepackage{hyperref} 
\usepackage{epsfig}
\usepackage[mathscr]{euscript}
\usepackage{amsmath}
\usepackage{graphicx}       
\usepackage{wrapfig}
\usepackage{dcolumn}  
\usepackage{bm}     
\usepackage{amssymb} 
\usepackage{titlesec}
\usepackage{mathtools}
\usepackage{relsize}
\usepackage{cleveref}
\usepackage{comment}
\usepackage{epsf}
\DeclareSymbolFont{boldoperators}{OT1}{cmr}{bx}{n}
\SetSymbolFont{boldoperators}{bold}{OT1}{cmr}{bx}{n}
\edef\bar{\unexpanded{\protect\mathaccentV{bar}}\number\symboldoperators16}
\begin{document}
\title{Driven synchronization in random networks of oscillators}
\author{Jason Hindes}
\affiliation{Laboratory of Atomic and Solid State Physics, Cornell University, Ithaca, New York}
\author{Christopher R. Myers}
\affiliation{Laboratory of Atomic and Solid State Physics, Cornell University, Ithaca, New York}
\affiliation{Institute of Biotechnology, Cornell University, Ithaca, New York}
\begin{abstract}
Synchronization is a universal phenomenon found in many non-equilibrium systems. Much recent interest in this area has overlapped with the study of complex networks, where a major focus is determining how a system's connectivity patterns affect the types of behavior that it can produce. Thus far, modeling efforts have focused on the tendency of networks of oscillators to mutually synchronize themselves, with less emphasis on the effects of external driving. In this work we discuss the interplay between mutual and driven synchronization in networks of phase oscillators of the Kuramoto type, and explore how the structure and emergence of such states depends on the underlying network topology for simple random networks with a given degree distribution. We find a variety of interesting dynamical behaviors, including bifurcations and bistability patterns that are qualitatively different for heterogeneous and homogeneous networks, and which are separated by a Takens-Bogdanov-Cusp singularity in the parameter region where the coupling strength between oscillators is weak. Our analysis is connected to the underlying dynamics of oscillator clusters for important states and transitions. 
\end{abstract}
\maketitle
{\quotation{Collective behavior of complex networks is a very active field of theoretical and practical research. In particular, models of oscillator networks have drawn much attention due to their numerous applications across diverse fields, with a particular emphasis on synchronization phenomena.  Here we study the dynamics of coupled oscillators, subject to periodic forcing, on random networks with different degrees of connectivity, and uncover many dynamical behaviors as a few parameters are varied. We find that the unfolding of synchronized states, and the possibility of bistability among them, differs for networks depending on how heterogeneous the degree of local connectivity is. This is explained through a combination of analytic and numerical results.}}
\section{\label{sec:Intro} INTRODUCTION}
The tendency for populations of oscillators to synchronize their dynamics and produce large-scale collective oscillations is relevant in a wide range of contexts \cite{Acebron, Balanov, Strogatz}. A particularly simple class of models for this behavior was proposed by Kuramoto, where each oscillator in a network is described by a phase variable, which has a tendency to oscillate at its natural frequency and in phase with its neighbors \cite{Kuramoto}. This model has given insights into the dynamics of many systems, from the synchronization of coupled chemical oscillators and Josephson junction arrays, to correlations in visual cortex experiments and coherence in neutrino flavor oscillations \cite{Sompolinsky,Kiss,Pantaleone,Wiesenfeld}. 

Much recent work on the Kuramoto model has concerned synchronization on complex networks, where the transition to coherent oscillations depends on the properties of the network topology \cite{Dorogovtsev, Arenas, Vespignani1}. Some important results are vanishing synchronization thresholds and explosive transitions for networks with large degree fluctuations \cite{Arenas, Gomez, Skardal}.  However, the effects of external driving are much less known, and questions about how different networks of oscillators respond to driving, and to what extent they can be controlled, have not been answered, even though in many circumstances, external fields are present \cite{Barabasi}. An important example is the network of pacemaker cells,  which play a role in determining mammalian circadian rhythms, and can be driven by light-dark cycles \cite{Wang,Dunlap,Liu}.    

In what follows, we discuss the \textit{interplay between mutual and driven synchronization} in random networks of phase oscillators with a given degree distribution. In particular, we present key aspects of the stability diagram for the driven Kuramoto model on these networks, focusing on the appearance of a codimension-three Takens-Bogdanov-Cusp singularity in the parameter region where the coupling strength between oscillators is weak. This bifurcation description is used to explain various pathways to driven and mutual synchronization in terms of synchronized oscillator clusters and network topology. 
  
\section{\label{sec:ModelReduction}MEAN-FIELD REDUCTION AND ANALYSIS}
Kuramoto showed that a system of limit cycle oscillators, each near their own Hopf bifurcation, with weak coupling to their neighbors and fast amplitude equilibration, have the following simple equations of motion: 
\begin{equation}
\label{eq:Kuramoto}
\frac{d\theta_{i}}{dt} = \omega_{i} + J\sum_{j}A_{ij}\sin(\theta_{j}-\theta_{i}),
\end{equation}
where $\theta_{i}$ is the phase of the $i$th oscillator, with natural frequency $\omega_{i}$, coupling strength $J$, and  adjacency matrix for the interaction network $A_{ij}$. Under generic circumstances (e.g., when the natural frequencies are randomly assigned according to a symmetric and unimodal distribution without correlations to the topology), this system undergoes a critical transition from incoherence to mutual synchronization once the coupling strength exceeds a threshold, resulting in a fraction of the network oscillating at the average of the natural frequencies, and with a stationary phase distribution \cite{Acebron,Strogatz,Skardal2}.

A simple extension of the Kuramoto model that includes a periodic driving force is given by \cite{Sakaguchi,Antonsen, Childs}:    
\begin{equation}
\label{eq:KuramotoPlusForcing}
\frac{d\theta_{i}}{dt} = \omega_{i} + J\sum_{j}A_{ij}\sin(\theta_{j}-\theta_{i}) + E\sin(\Omega t -\theta_{i}),
\end{equation}
with external field strength $E$ and frequency $\Omega$. With similar assumptions for the natural frequency distribution, we expect each term to have the following effects on the dynamics: the randomness in the frequencies causes oscillators to have disperse phases with monotonic build-up in time, the coupling tends to align the phases of neighbors in proportion to the number of connections in a local environment (which will vary across the network), and the driving field tends to force oscillators to move at the driving frequency and away from its natural frequency. The interactions among these tendencies, both cooperative and competitive, will depend on the magnitude of each term and the network topology, and therefore we expect an intricate dynamics with multiple behaviors and transitions \cite{Childs}. 
\subsection{\label{sec:DegreeClassDynamics}  Degree class dynamics}
To clarify the dynamics, we attempt to find a reduced description of \eqref{eq:KuramotoPlusForcing}. For convenience, we study the phases in the co-moving frame of the driving, $\phi_{i}=\theta_{i}-\Omega t$:
\begin{equation}
\label{eq:ForcingCMFram}
\frac{d\phi_{i}}{dt} = \omega_{i}-\Omega + J\sum_{j}A_{ij}\sin(\phi_{j}-\phi_{i}) - E\sin(\phi_{i}),
\end{equation} 
and consider random networks with a given degree distribution, $p_{k}$, that specifies the fraction of oscillators with $k$ neighbors. In particular, we will study the annealed limit of random networks explicitly, for which $A_{ij}= \tfrac{k_{i}k_{j}}{N\left<k\right>}$, where $k_{i}$ and $k_{j}$ are drawn from $p_{k}$ for a network of size $N$ with average degree $\left<k\right>$, but note that our results are in qualitative agreement with quenched models (such as the configuration model, Fig.\ref{fig:OP}). For annealed networks we find oscillator dynamics:  
\begin{equation}
\label{eq:AnnealedDynamics}
\frac{d\phi_{i}}{dt}=\omega_{i}-\Omega+J k_{i} \mathcal{I}m\!\left[\!e^{-i\phi_{i}}\sum_{j}\frac{k_{j}e^{i\phi_{j}}}{\!N\!\left<k\right>}\!\right] +E \mathcal{I}m\!\left[e^{-i\phi_{i}}\!\right],
\end{equation}   
from which we can define the complex order parameter
\begin{equation}
\label{eq:OderParamFiniteN}
z= \sum_{j}\frac{k_{j}e^{i\phi_{j}}}{\!N\left<k\right>},
\end{equation}    
or the average interaction strength (both magnitude and phase) that an oscillator feels along an edge to its neighbors. 

We are interested in the thermodynamic limit, $N \to \infty$, in which it is useful to consider the density of oscillators with phase $\phi$ at time $t$, given degree $k$ and frequency $\omega$, $\rho(\phi , t ; k,\omega)$. This probability density satisfies a continuity relation: 
\begin{equation}
\label{eq:Continuity}
\frac{\partial \rho}{\partial t}\!=-\frac{\partial}{\partial \phi}\!\bigg[\!\Big(\omega-\Omega + \frac{e^{{-}i\phi}}{2i}\!\left(Jkz+E\right) - \frac{e^{i\phi}}{2i}\!\left(Jk\bar{z}+E\right)\!                                               \!\Big)\rho\bigg]
\end{equation}
with 
\begin{equation}
\label{eq:OderParameterIntegral}
z(t)= \sum_{k}\frac{kp_{k}}{\left<k\right>}\int g(\omega) \rho(\phi,t;\omega,k) e^{i\phi} d\omega d\phi,
\end{equation}
where $g(\omega)$ is the natural frequency distribution and $\bar{z}$ is the complex conjugate of $z$. In order to solve \eqref{eq:Continuity} we expand $\rho$ into its Fourier components:
\begin{align}
\label{eq:FourierSeries}
\rho(\phi,t; \omega,k)&= \frac{1}{2\pi}\Big[1+\sum_{n=1}^{\infty}\alpha_{n}(t; \omega,k)e^{in\phi} +\text{c.c.} \Big],
\end{align}
and look for simple power-series solutions of the form, $\alpha_{n}(t;\omega,k)= \alpha^{n}(t;\omega,k)$ -- an ansatz which was proposed by Ott and Antonsen, and that is applicable in a wide array of Kuramoto model variants \cite{Ott,Pazo,Ji}. In this case it gives the dynamics for $\bar{\alpha}(t;\omega,k)$: 
\begin{equation}
\label{eq:FirstHarmonic}
\frac{d\bar{\alpha}}{dt} = \frac{1}{2}\big[Jkz+E\big]+i\left(\omega-\Omega\right)\bar{\alpha}-\frac{1}{2}\big[Jk\bar{z}+E\big]\bar{\alpha}^2 , 
\end{equation} 
\noindent which completely specifies the order parameter:  
\begin{equation}
\label{eq:OderParameterIntegralReduced}
z(t)= \sum_{k}\frac{kp_{k}}{\left<k\right>}\int g(\omega) \bar{\alpha}(t;\omega,k) d\omega . 
\end{equation}
In addition, the dimensionality of the system can be further reduced by performing the natural frequency integral, for which we assume: 
\begin{equation}
\label{eq:FreqDist}
g(\omega)= \frac{\gamma}{\pi\Big[\left(\omega-\omega_{0}\right)^2+\gamma^2\Big]},
\end{equation}
\noindent a Cauchy distribution with median $\omega_{0}$ and scale $\gamma$. Generically, $\bar{\alpha}(\omega,k,t)$ has no poles in the upper-half of the complex $\omega$-plane, and therefore we perform contour integration of \eqref{eq:OderParameterIntegralReduced} closed in this region \cite{Ott}, which reduces the integral to the residue at the pole $\omega_{0}+i\gamma$: 
\begin{equation}
\label{eq:OderParameterFinal}
z(t)= \sum_{k}\frac{kp_{k}}{\left<k\right>} \bar{\alpha}(t;\omega_{0}+i\gamma,k) \equiv  \sum_{k}\frac{kp_{k}}{\left<k\right>} a_{k}(t), 
\end{equation}
\noindent where 
\begin{equation}
\label{eq:Fundamental}
\frac{da_k}{d\tau} = \frac{1}{2}\big[\mathcal{J} kz+\mathcal{E} \big]-\left(1+i\Delta\right)a_{k}-\frac{1}{2}\big[\mathcal{J}k\bar{z}+\mathcal{E} \big]a_{k}^{2} ,
\end{equation} 
with the dimensionless time, $\tau=\gamma t$, and normalized parameters: $\mathcal{E}=E/\gamma$, $\mathcal{J}=J/\gamma$, and $\Delta= (\Omega-\omega_{0})/\gamma$. 

This is the fundamental equation for the thermodynamic limit of the forced Kuramoto model on annealed networks. The dynamics has been reduced to a description of the average contribution to the order parameter by nodes of degree $k$, with the size of the state-space equal to twice the number of degree classes. In the following, we will focus on networks that have power-law degree distributions with finite cutoffs and Poisson distributions, 
\vspace{-0.3cm}
\begin{equation}
\label{eq:DegreeDist}
p_{k}= \frac{k^{-s}}{\sum\limits_{k^{'}=1}^{K_{cut}}{k^{'}}^{-s}} \;\;\;\; \text{and} \;\;\;\;  p_{k}=\frac{{e^{ - {\left<k\right>} } {\left<k\right>} ^k }}{{k!}},  
\end{equation} 

\noindent respectively, though analytic results are given for arbitrary distributions. We will refer to the former as simply ``power-law'' for brevity, though the degree cutoff, $K_{cut}$, will be specified when pertinent. In general, the cutoff determines the dimensionality of the reduced system, and we find that its value is relevant for heterogeneous network behavior, where large degree nodes can contribute significantly to the dynamics.   

\subsection{\label{sec:Bifurcations} Limiting states}
First, we consider the states of mutual and driven synchronization in instructive limits.  For instance, in the limit where $\mathcal{E}\rightarrow0$, Eq.\eqref{eq:Fundamental} describes an un-driven network,  and has stable solutions corresponding to oscillating waves: $a_{k}=r_{k}(\tau)e^{-i\Delta \tau}$, $z=R(\tau)e^{-i\Delta \tau}$,  
\begin{equation}
\label{eq:UnDrivenLimit}
\frac{dr_k}{d\tau} = \frac{1}{2}\mathcal{J}kR \big[1-r_{k}^2]-r_{k},
\end{equation} 
which reproduces known results \cite{Ji}. In particular for the frame where $\Omega=0$, the network tends to a purely oscillating state at the average natural frequency $\omega_{0}$, with some fixed ${r^{*}_{k}}$. In addition, the incoherent state, $r_{k}^{*}=0$, has a linear stability exponent 
\begin{equation}
\label{eq:Incoherent}
 \lambda_{ic}=\frac{\mathcal{J}\left<k^{2}\right>}{2\left<k\right>}-1, 
\end{equation} 
which implies a threshold for the onset of mutual synchronization in the absence of driving, $r_{k}^{*}\neq0$,  $\frac{\mathcal{J}_{c}\left<k^{2}\right>}{2\left<k\right>}=1$. We consider situations where $\mathcal{J} >\mathcal{J}_{c}$, and a coherent mutually synchronized state is stable without forcing \cite{Arenas,F1}. 

On the other hand, in the limit where the driving frequency is equal to the average natural frequency, $\Delta\rightarrow0$, Eq.\eqref{eq:Fundamental} describes states of driven synchronization, where the network is oscillating at the driving frequency on average, with amplitudes given by the fixed points of the self-consistent equation: 
\begin{equation}
\label{eq:DrivenLimit}
R^{*}= \sum_{k}\frac{kp_{k}}{\left<k\right>} \Bigg(\frac{-1+\sqrt{1+\big(\mathcal{J}kR^{*}+\mathcal{E}\big)^2}}{\big(\mathcal{J}kR^{*}+\mathcal{E}\big)}\Bigg),
\end{equation} 
and with a large number of nodes entrained to the driving. From \eqref{eq:DrivenLimit} we can see that incoherence is not a solution when $\mathcal{E}\neq0$, meaning that external driving always enforces some level of coherent oscillations at its frequency. Moreover, multiple coherent solutions can exist depending on the parameter values.  

\subsection{\label{sec:StabilityDiagram} Partial stability diagram} 
Next, we provide the results of a stability and partial\cite{F2} bifurcation analysis that delineates the boundaries between the limiting states and helps to explain how each can be converted into the other. The associated stability diagrams are somewhat complicated, and it is therefore useful to have the results in hand before proceeding to fill in the details. A quantitative discussion and analysis can be found in Sec.\ref{sec:BifurcationAnalysis} that derives some of the results, with a broader summary of behaviors found in Sec.\ref{sec:KeyTransitions}.\\

\begin{figure}[H]
\raggedleft{\includegraphics[scale=0.189]{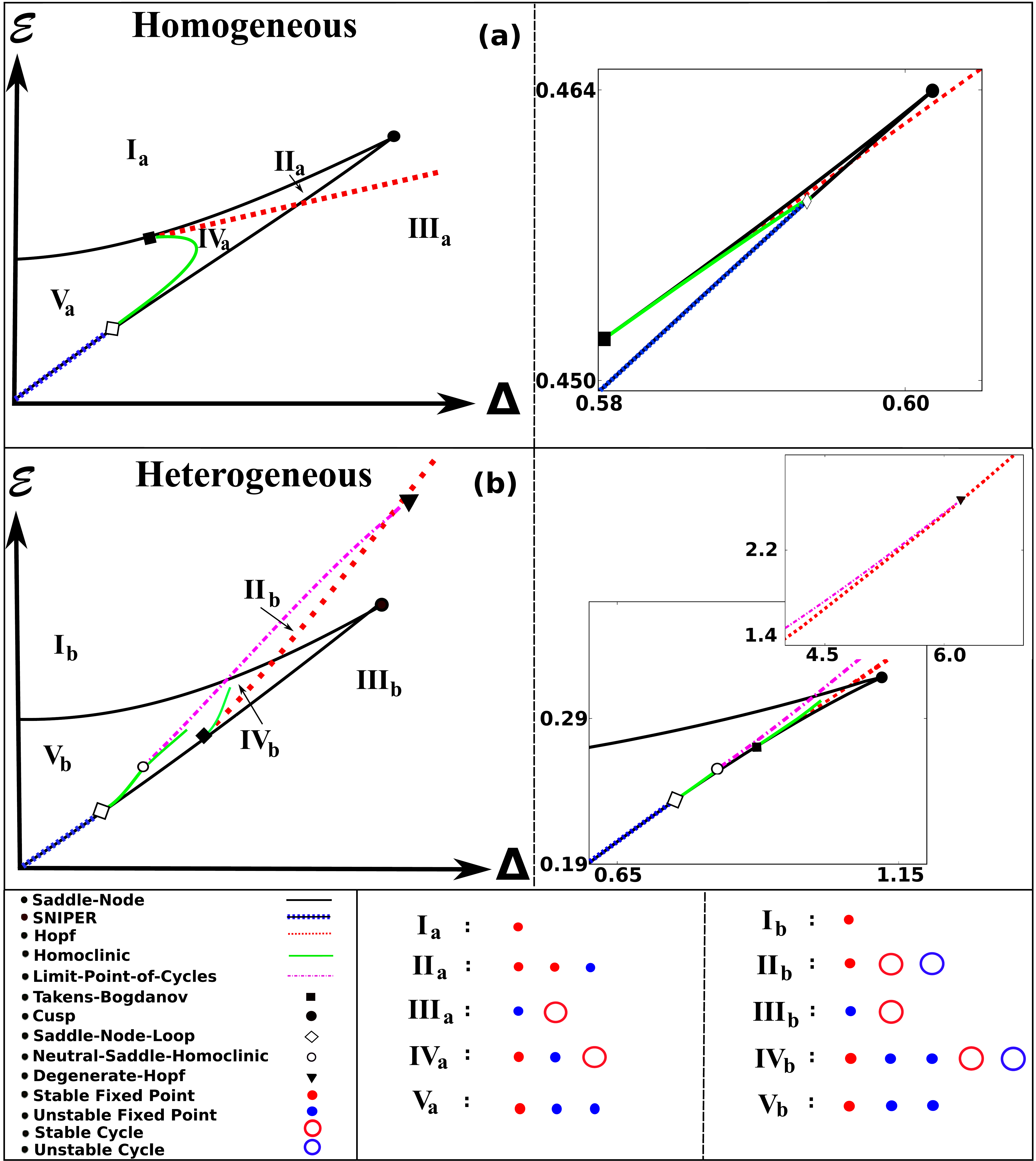}}
\caption{{{(Color online) 
Stability diagrams for the driven Kuramoto model on random networks shown as functions of the driving field strength $\mathcal{E}$ and frequency detuning $\Delta$. A legend is given in the bottom left. (a) Schematic diagram (left) for homogeneous behavior (e.g., power-laws with large exponents,  k-regular, and Poisson degree distributions); (right) diagram for a power-law network with $\mathcal{J}=2$, $s=3.0$, and $K_{cut}=200$. (b) Schematic partial\cite{F2} diagram (left) for heterogeneous behavior (power-law degree distribution with small exponent), in the parameter region where $\mathcal{J}$ is weak (Sec.\ref{sec:BifurcationAnalysis}); partial diagram for a power-law network with $\mathcal{J}=0.25$, $s=2.3$, and $K_{cut}=1000$, shown in two parameter ranges for clarity. A table indicating possible states is shown in the bottom right for important regions (Roman numerals).   
}}} 
\label{fig:Bifurcation} 
\end{figure} 

The schematic stability diagrams shown in Fig.\ref{fig:Bifurcation} illustrate two types of behavior in the $(\Delta, \mathcal{E})$ plane when the coupling, $\mathcal{J}$, is weak (Sec.\ref{sec:BifurcationAnalysis}). If we  consider networks with power-law degree distributions, Fig.\ref{fig:Bifurcation}(a) shows the generic behavior when the degree exponent, $s$, is large. We find that this is maintained for networks with relatively homogeneous degree distributions, such as Erd\H{o}s R\'{e}nyi, k-regular, or complete graphs \cite{Childs}. Conversely when the degree exponent is small, i.e., the degree distribution has a heavy tail, the behavior looks like Fig.\ref{fig:Bifurcation}(b). Because the former reproduces the behavior for the complete graph, and the latter occurs as the amount of variation in the degree distribution is increased, we distinguish these cases by the terms homogeneous and heterogeneous driven behavior. 

In particular, we find that the Takens-Bogdanov point appears on the upper branch of the saddle-node bifurcations for the homogeneous case, but appears on the lower branch for the heterogenous case, as depicted in Fig.\ref{fig:Bifurcation}. The transition between the two behaviors can occur, for example, by decreasing the degree exponent, for some fixed $\mathcal{J}$ and $K_{cut}$, until the Takens-Bogdanov and cusp bifurcations are coincident, which typically occurs for some $2\lesssim s \lesssim3$ (see Sec.\ref{sec:BifurcationAnalysis} for descriptions of these bifurcations). The existence of this singularity allows us to construct the behaviors shown through a combination of analytic results, numerical continuation, and general predictions for subsequent bifurcations. Details are given in the following sections, and example stability diagrams are shown alongside schematics in Fig.\ref{fig:Bifurcation}.      

\subsection{\label{sec:BifurcationAnalysis} Stability analysis and bifurcations}  
We begin constructing the stability digrams by first finding the fixed points of Eq.\eqref{eq:Fundamental}, which denote states of driven synchronization, and establish how such states change stability. In general, fixed points satisfy $\frac{da_{k}}{d\tau}=0$, which implies the self-consistent condition for $z^{*}$:
\begin{align}
\label{eq:FixedPoint}
z^{*}=& \sum_{k}\frac{kp_{k}}{\left<k\right>}a_{k}^{*} \nonumber \\
=&\sum_{k}\frac{kp_{k}}{\left<k\right>}\frac{-\left(1+i\Delta\right)+\sqrt{\left(1+i\Delta\right)^2+|\mathcal{J} kz^{*}+\mathcal{E}|^{2}}}{\mathcal{J} k\bar{z}^{*}+\mathcal{E}}. 
\end{align}

Each $a_{k}$ is a complex number, and so could be represented by a magnitude and phase, or with real and imaginary parts. Next, it is useful to consider how the dynamics respond to perturbations away from the steady-states given by \eqref{eq:FixedPoint}, e.g. $\mathcal{R}e[a_{k}^{*}]+x_{k}$ and  $\mathcal{I}m[a_{k}^{*}]+y_{k}$ where $x_{k}$ and $y_{k}$ are the $k$'th components of the right eigenvectors of \eqref{eq:Fundamental} at $a_{k}^{*}$, in the real and imaginary part representation of $a_{k}$. Equivalently, we can define $\eta_{k}=\frac{1}{\sqrt{2}}(x_{k}+i y_{k})$ and $\widetilde{\eta_{k}}=\frac{1}{\sqrt{2}}(x_{k}-iy_{k})$, with the perturbations $a_{k}^{*}+\sqrt{2}\eta_{k}$ and $\bar{a}_{k}^{*}+\sqrt{2}\widetilde{\eta_{k}}$. It is more convenient to use the latter and leave Eq.\eqref{eq:Fundamental} in its complex form, while keeping in mind that the standard results of bifurcation theory pertain to some underlying real representation of \eqref{eq:Fundamental}. 

We look for the linear stability spectrum of eigen-modes around a fixed point by adding the perturbations discussed into \eqref{eq:Fundamental}, and collecting terms of order $\eta$:
\begin{align}
\label{eq:Perturbation}
\frac{d\eta_{k}}{d\tau} &= \frac{\mathcal{J} k}{2}\Bigg[\sum_{k'}\frac{k'p_{k'}}{\left<k\right>}\eta_{k'} - a_{k}^{\boldmath{*}^\mathlarger{2}}\sum_{k'}\frac{k'p_{k'}}{\left<k\right>}\widetilde{\eta}_{k'} \Bigg]-q_{k}^{*}\eta_{k}, \nonumber \\
\frac{d\widetilde{\eta}_{k}}{d\tau} &= \frac{\mathcal{J} k}{2}\Bigg[\sum_{k'}\frac{k'p_{k'}}{\left<k\right>}\widetilde{\eta}_{k'} - \bar{a}_{k}^{\boldmath{*}^\mathlarger{2}}\sum_{k'}\frac{k'p_{k'}}{\left<k\right>}\eta_{k'} \Bigg]-\bar{q}_{k}^{*}\widetilde{\eta}_{k}, \nonumber \\
\end{align} 
with 
\begin{align}
q_{k}^{*}&=1+i\Delta +\left(\mathcal{J} k \bar{z}^{*} +\mathcal{E}\right)a_{k}^{*};
\end{align} 
This system has a set of solutions, $\frac{d\eta_{k}}{d\tau}=\lambda \eta_{k}$ and $\frac{d\widetilde{\eta}_{k}}{d\tau}=\lambda \widetilde{\eta}_{k}$, from which we can find a self-consistent equation for the spectrum $\{\lambda\}$. Solving for $\eta_{k}$ and $\widetilde{\eta_{k}}$ in \eqref{eq:Perturbation}, multiplying by $\frac{kp_{k}}{\left<k\right>}$, summing over $k$, and eliminating the constants $\sum_{k}\frac{kp_{k}}{\left<k\right> }\eta_{k}$  and $\sum_{k}\frac{kp_{k}}{\left<k\right>}\widetilde{\eta}_{k}$ gives:
 \begin{align}
\label{eq:Spectrum}
\Biggr[{   \frac{\sum\limits_{k}\frac{\mathcal{J}k^{2}p_{k}a_{k}^{\boldmath{*}^\mathlarger{2}}}{2\left<k\right>(\lambda+q^{*}_{k})}}{\sum\limits_{k}\frac{\mathcal{J}k^{2}p_{k}}{2\left<k\right>(\lambda+q^{*}_{k})} -1} \Biggr]  \Biggr[   \frac{\sum\limits_{k}\frac{\mathcal{J}k^{2}p_{k}\bar{a}_{k}^{\boldmath{*}^\mathlarger{2}}}{2\left<k\right>(\lambda+\bar{q}^{*}_{k})}}{\sum\limits_{k}\frac{\mathcal{J}k^{2}p_{k}}{2\left<k\right>(\lambda+\bar{q}^{*}_{k})} -1} }\Biggr]=1. 
\end{align} 

Next, we catalogue relevant bifurcations found in Fig.\ref{fig:Bifurcation}, and discuss their dynamical behaviors in Sec. \ref{sec:KeyTransitions}. First, the spectrum condition can be used to find the local codimension-one bifurcations, where some number of eigenvalues cross the imaginary axis (codimension implying the number of parameters that must be changed in order for a bifurcation to occur) \cite{Crawford, Strogatz2}. The most generic such crossing is the saddle-node bifurcation ($SN$), in which the spectrum at the equilibrium has one simple zero eigenvalue, and at which two equilibrium points collide and disappear: 
\begin{equation}
\label{eq:SaddleNode}
\left|\Biggr[ \frac{\sum\limits_{k}\frac{\mathcal{J}k^{2}p_{k}a_{k}^{\boldmath{*}^\mathlarger{2}}}{2\left<k\right>q^{*}_{k}}}{\sum\limits_{k}\frac{\mathcal{J}k^{2}p_{k}}{2\left<k\right>q^{*}_{k}} -1} \Biggr]\right|^2 =1. 
\end{equation}
The $SN$ condition \eqref{eq:SaddleNode} predicts when steady states of driven synchronization vanish, and signifies when a local barrier (represented by the saddle) in the dynamics has been overcome. Importantly, we find that the lower branch of $SN$ bifurcations contains a section of saddle-node-infinite-period bifurcations ($SNIPER$) (e.g., crossing V-III in Fig.\ref{fig:Bifurcation}), where an $SN$ occurs on a limit cycle of infinite period \cite{Childs}.

Another local codimension-one bifurcation is the Hopf ($H$), in which the spectrum at the equilibrium has two purely imaginary eigenvalues, with all others having non-zero real parts. At this point the amplitude of a periodic orbit decreases continuously to zero with its period tending to $2\pi/\omega_{H}$, where $\lambda=i\omega_{H}$:
\begin{equation}
\label{eq:Hopf}
\Biggr[   \frac{\sum\limits_{k}\frac{\mathcal{J}k^{2}p_{k}a_{k}^{\boldmath{*}^\mathlarger{2}}}{2\left<k\right>(i\omega_{H}+q^{*}_{k})}}{\sum\limits_{k}\frac{\mathcal{J}k^{2}p_{k}}{2\left<k\right>(i\omega_{H}+q^{*}_{k})} -1} \Biggr]  \Biggr[   \frac{\sum\limits_{k}\frac{\mathcal{J}k^{2}p_{k}\bar{a}_{k}^{\boldmath{*}^\mathlarger{2}}}{2\left<k\right>(i\omega_{H}+\bar{q}^{*}_{k})}}{\sum\limits_{k}\frac{\mathcal{J}k^{2}p_{k}}{2\left<k\right>(i\omega_{H}+\bar{q}^{*}_{k})} -1} \Biggr]  =1. 
\end{equation}
When the periodic orbit associated with the Hopf bifurcation is stable, it is called supercritical ($H_{sup}$), and when it is unstable, it is called subcritical $(H_{sub})\;\;$ \cite{Strogatz2, Kuznetsov1}. In contrast with homogenous network behavior (e.g., crossing $\text{I}_{a}-\text{III}_{a}$ in Fig.\ref{fig:Bifurcation}(a))$\;$ \cite{Childs}, both branches of cycle-stability can appear if the degree distribution is broad enough (e.g., crossing $\text{II}_{b}-\text{III}_{b}$ in Fig.\ref{fig:Bifurcation}(b)).

Beyond the local codimension-one bifurcations, there are two key local codimension-two bifurcations. These are important to unravel because they can inform us as to what global bifurcations occur. The first appears when two branches of the $SN$ collide, in the neighborhood of which there exist three states of driven entrainment; this is known as a cusp $(C)\;$ \cite{Kuznetsov1}. To find the $C$ point, we first consider that near a bifurcation, the equations of motion can be restricted to a center manifold with the same dimension as the number of eigenvectors whose eigenvalues cross the imaginary axis, and is tangent to those vectors. Furthermore, the dynamics of the center manifold are equivalent to the normal form for the bifurcation. In the simple case of a $SN$, the center manifold is one-dimensional, $m=w\eta+w^{2}h+\mathcal{O}(w^{3})$ with the normal form: $\frac{dw}{dt}=cw^{2}+\mathcal{O}(w^{3})\;\;$ \cite{Crawford, Kuznetsov1}. 

The $C$ bifurcation occurs when $c=0$, a  condition for which can be found by substituting the center manifold expansion and normal form into \eqref{eq:Fundamental}, collecting terms of order $w^{2}$, and taking the complex inner product of the resulting vector, $B$, with the left eigenvectors, $\zeta\;\;$ \cite{Kuznetsov2}. The right and left eigenvectors are found from a similar self-consistent analysis as for \eqref{eq:Spectrum}, and in the complex representation are respectively:
\begin{align}
\label{eq:EigenVectors}
\eta_{k}(\lambda)&=\frac{A\mathcal{J}k}{2}\Bigg(\frac{x(\lambda)-a_{k}^{\boldmath{*}^\mathlarger{2}}}{\lambda+q^{*}_{k}}\Bigg)\\ 
\widetilde{\eta}_{k}(\lambda)&=\frac{A\mathcal{J}k}{2}\Bigg(\frac{1-x(\lambda)\bar{a}_{k}^{\boldmath{*}^\mathlarger{2}}}{\lambda+\bar{q}^{*}_{k}}\Bigg)\\
\zeta_{k}(\lambda)&=\frac{Zkp_{k}}{\left<k\right>}\Biggr(\frac{x(\lambda)}{\Big(\sum\limits_{k}{\frac{\mathcal{J}k^{2}p_{k}}{2\left<k\right>(\lambda+\bar{q}^{*}_{k})} -1} \Big)\Big(\lambda+\bar{q}^{*}_{k}\Big)}\Biggr)\\ 
\widetilde{\zeta}_{k}(\lambda)&=\frac{Zkp_{k}}{\left<k\right>}\Biggr(\frac{1}{\Big(\sum\limits_{k}{\frac{\mathcal{J}k^{2}p_{k}}{2\left<k\right>(\lambda+q^{*}_{k})} -1} \Big)\Big(\lambda+q^{*}_{k}\Big)}\Biggr),
\end{align}
with constants $A$ and $Z$, and with the conveniently defined sum, 
\begin{equation}
\label{eq:Sum}
x(\lambda)=\Biggr[\frac{\sum\limits_{k}\frac{\mathcal{J}k^{2}p_{k}a_{k}^{\boldmath{*}^\mathlarger{2}}}{2\left<k\right>(\lambda+q^{*}_{k})}}{\sum\limits_{k}\frac{\mathcal{J}k^{2}p_{k}}{2\left<k\right>(\lambda+q^{*}_{k})} -1} \Biggr].  
\end{equation} 
Collecting terms of order $w^{2}$ in the expansion produces the bilinear form for \eqref{eq:Fundamental} evaluated at the vector $\eta_{k}, \widetilde{\eta}_{k}$: 
\begin{align}
\label{eq:Bilinear}
B_{k}(\lambda)&=-2\mathcal{J}Ak\eta_{k}a_{k}^{*}-(\mathcal{J}k\bar{z}^{*}+\mathcal{E})\eta_{k}^{2}\\
\widetilde{B}_{k}(\lambda)&=-2\mathcal{J}Axk\widetilde{\eta}_{k}\bar{a}_{k}^{*}-(\mathcal{J}kz^{*}+\mathcal{E})\widetilde{\eta}_{k}^{2}.
\end{align}
Putting these together generates the normal form coefficient $c$, and a condition for the cusp bifurcation:
\begin{equation}
\label{eq:Cusp}
c=\sum_{k} \bar{\zeta}_{k}(0)B_{k}(0)+ \bar{\widetilde{\zeta}}_{k}(0)\widetilde{B}_{k}(0)=0,
 \end{equation}  
 in conjunction with (20).

It should be noted that for power-law networks with $\mathcal{J} \lesssim 2.5$, the number of possible fixed points for this system is three, which we call the weak coupling region. However, when the coupling is stronger, a degenerate $C$ point seems to emerge, which generates additional unstable and saddle states, and complicates the unfolding (shown in Fig.\ref{fig:Bifurcation}), though much of the general structure is maintained for larger $\mathcal{J}$. In this work, we restrict ourselves to the weak coupling region for power-law networks, because the comparison between homogeneous and heterogeneous graphs is more straightforward. 

The second local codimension-two bifurcation is the Takens-Bogdanov ($TB$), at which the spectrum has a double root at zero. Attached to this bifurcation are curves of $SN$ and $H$ bifurcations as well as a curve of Homoclinic Bifurcations $(HC)\;$ \cite{Kuznetsov1, Kuznetsov2, Kuznetsov3, Danglemayr, Guckenheimer}. In the latter, the period of a cycle diverges as it collides with a saddle-point and connects its stable and unstable manifolds (e.g., crossing $\text{IV}_{a}-\text{V}_{a}$ in Fig.\ref{fig:Bifurcation}(a)). To find the location of the $TB$ bifurcation, we expand \eqref{eq:Hopf} in powers of $\omega_{H}$, and enforce that terms of order $\omega_{H}$ vanish, which gives the criterion: 
\begin{multline}
\label{eq:TB}
\mathcal{R}e\Biggr[ \Bigg(\sum\limits_{k}\frac{\mathcal{J}k^{2}p_{k}a_{k}^{\boldmath{*}^\mathlarger{2}}}{2\left<k\right>{q^{*}_{k}}^{2}} \Bigg)\Bigg(\sum\limits_{k}\frac{\mathcal{J}k^{2}p_{k}\bar{a}_{k}^{\boldmath{*}^\mathlarger{2}}}{2\left<k\right>\bar{q}^{*}_{k}} \Bigg) - \\
\Bigg(\sum\limits_{k}\frac{\mathcal{J}k^{2}p_{k}}{2\left<k\right>{q^{*}_{k}}^{2}} \Bigg)\Bigg(\sum\limits_{k}\frac{\mathcal{J}k^{2}p_{k}}{2\left<k\right>\bar{q}^{*}_{k}} -1 \Bigg) \Biggr] =0,  
\end{multline}
that in conjunction with \eqref{eq:SaddleNode}, determines the bifurcation point.   

Finally, the highest codimension bifurcation that we consider arises when the $C$ collides with the $TB$ ($TBC$), implying that \eqref{eq:SaddleNode}, \eqref{eq:Cusp}, and \eqref{eq:TB} are all satisfied (which also occurs in the Hodgkin-Huxley equations) \cite{Kuznetsov2, Guckenheimer, Mohieddine}. In addition to the bifurcations discussed, this particular singularity predicts curves of codimension-two homoclinic bifurcations to Saddle-Node-Loops ($SNL$) and Neutral Saddles ($NS$), and curves of Degenerate Hopf bifurcations ($DH$). The latter two are termination points for curves of Limit-Point-of-Cycles ($LPC$). These bifurcations imply new  
behaviors that do not appear for homogeneous networks and have interesting effects on the dynamics. Specifically, the $LPC$ transition entails that a stable cycle collides with an unstable cycle and disappears, while the $DH$ entails that an $LPC$ emerges on a $H$ point -- typically as the Lyapunov exponent of the Hopf cycle vanishes. Because these bifurcations only occur when the $TB$ is on the lower branch of the $SN$, they are not seen in homogeneous networks. Lastly, the $SNL$ and $NS$ entail that a homoclinic cycle is coincident with a $SN$ and a saddle whose whose eigenvalues sum to zero, respectively \cite{Kuznetsov1, Kuznetsov2, Guckenheimer,Danglemayr, Mohieddine} (see Fig.\ref{fig:Bifurcation}). 

\section{\label{sec:OscillatorDynamics}OSCILLATOR DYNAMICS} 
In this section, we explore some of the implications of the behaviors discussed on the dynamics of driven homogeneous and heterogeneous networks. Both the mean-field (Eq.\ref{eq:Fundamental}) and oscillator (Eq.\ref{eq:ForcingCMFram}) dynamics are examined.  
\subsection{\label{sec:KeyTransitions} Key transitions and bistability}    
\begin{figure}[t]
\raggedright{\includegraphics[scale=0.475]{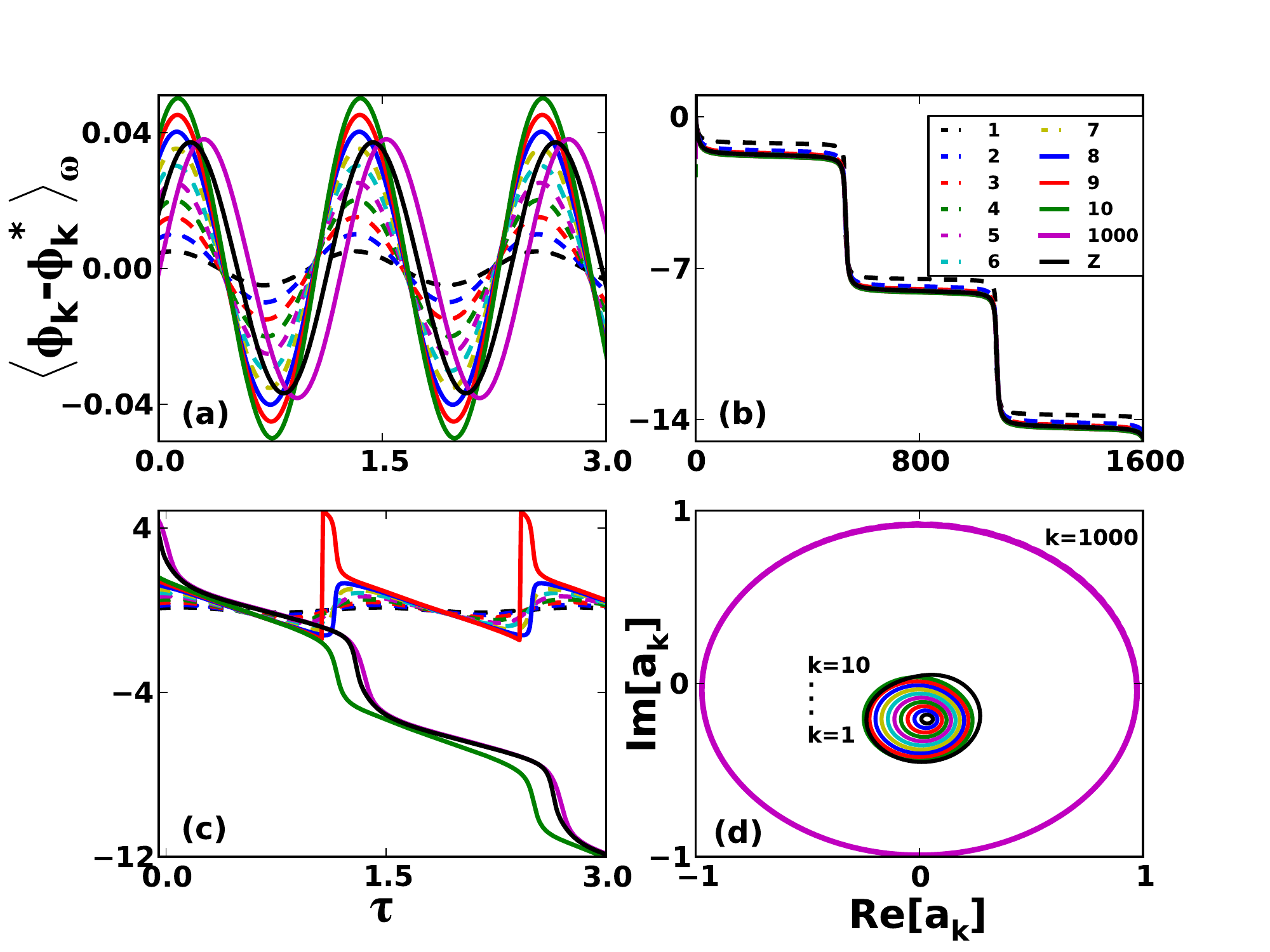}}
\caption{{{(Color online) Comparison of mutually synchronized states that arise from perturbations to driven states just below the key transitions for networks with power-law degree distributions. Subplots (a-c) show the average phase deflection from a driven state for various degree classes versus time; colors for every degree class are specified in (b). (a) Below a $H_{sup}$: $\mathcal{E}=\mathcal{E}_{H}-10^{-4}$, $\Delta=5.5$, $\mathcal{J}=0.25$, $s=2.6$, and $K_{cut}=1000$; the state appears with finite frequency and small amplitude. (b) Below a $SNIPER$: $\mathcal{E}=\mathcal{E}_{SN}-10^{-4}$, $\Delta=0.2$, $\mathcal{J}=0.25$, $s=2.3$, and $K_{cut}=1000$; the state appears with a large period and with all degree classes increasing phase monotonically with respect to the field. (c) Below a $LPC$ (below $H_{sub}$): $\mathcal{E}=\mathcal{E}_{H}-10^{-4}$, $\Delta=5.5$, $\mathcal{J}=0.25$, $s=2.3$, and $K_{cut}=1000$; the state appears with finite frequency and large amplitude, and with high degree nodes increasing phase monotonically with respect to the field (phase-slip motion), while small degree nodes remain phase-trapped (on average). (d) Mutually synchronized state of \eqref{eq:Fundamental} with (c) parameter values, illustrating the cycle size variation with degree for states produced by crossing the $LPC$ transition.}}}  
\label{fig:MutualComp} 
\end{figure}

First, we can distill from the above that there are three primary ways that a stable mutually synchronized state can be created: $H_{sup}$, $SNIPER$, and $LPC$ transitions. 
Qualitatively, we can think of such states as limit cycles, and can consider how their average amplitude and frequency (inverse period) emerge through each transition. If we imagine changing one parameter (e.g. $\mathcal{E}$), one of three things happens: the amplitude can appear continuously with a discontinuous frequency ($H_{sup}$), the amplitude can appear discontinuously with continuous frequency ($SNIPER$), or the amplitude and frequency can both appear discontinuously ($LPC$). The special case of continuous amplitude and frequency appearance occurs through a $TB$ bifurcation. Fig.\ref{fig:MutualComp} shows a comparison between the behaviors of mutually synchronized states produced by crossing these transitions.

Interestingly, we find that each transition has a signature in the average phase build-up with respect to the driving field. For example if the $SNIPER$ transition is crossed (e.g., crossing $\text{V}-\text{III}$ in Fig.\ref{fig:Bifurcation}), the order-parameter dynamics is a large limit cycle that includes the origin \cite{Childs}. This implies that the average phase of the network grows monotonically with respect to the field, and is therefore largely de-pinned from it, with a macroscopic number of nodes lapping it continually. Moreover, this behavior holds widely for degree classes as well -- most degrees continually lap the field on average, perhaps excluding low degree nodes (e.g., k=1 or 2) depending on the parameters (Fig.\ref{fig:MutualComp}(b)). On the other hand if the $H_{sup}$ is crossed (e.g., crossing $\text{I}-\text{III}$ in Fig.\ref{fig:Bifurcation}), a small limit cycle emerges, centered around an unstable driven state. In this case there is no net build-up of the average phase with respect to the field; the motion is analogous to quasi-periodicity with average frequency equal to the driving, and an emergent ``wobble" frequency given by \eqref{eq:Hopf}$\;$ \cite{Childs}. This behavior holds for all degree classes, implying that large and small degrees on average both have phase-trapped dynamics (Fig.\ref{fig:MutualComp} (a)). However if the LPC transition is crossed (e.g., crossing $\text{I}_{b}-\text{II}_{b}$ in Fig.\ref{fig:Bifurcation}(b)), a large cycle emerges for the order-parameter that includes the origin (similar to the SNIPER), but only holds for nodes with large degree on average, i.e., nodes of small degree undergo phase-trapped motion, while nodes of large degree undergo phase-slip motion (Fig.\ref{fig:MutualComp}(c)). If we consider moving up the $LPC$ by increasing $\mathcal{E}$, more and more high degree nodes become trapped by the field, until all are trapped, and the $H_{sup}$ occurs -- the opposite limit brings us to the lower $SNIPER$ (see Fig.\ref{fig:Bifurcation}).  

\begin{figure}[H]
\raggedleft{\includegraphics[scale=0.25]{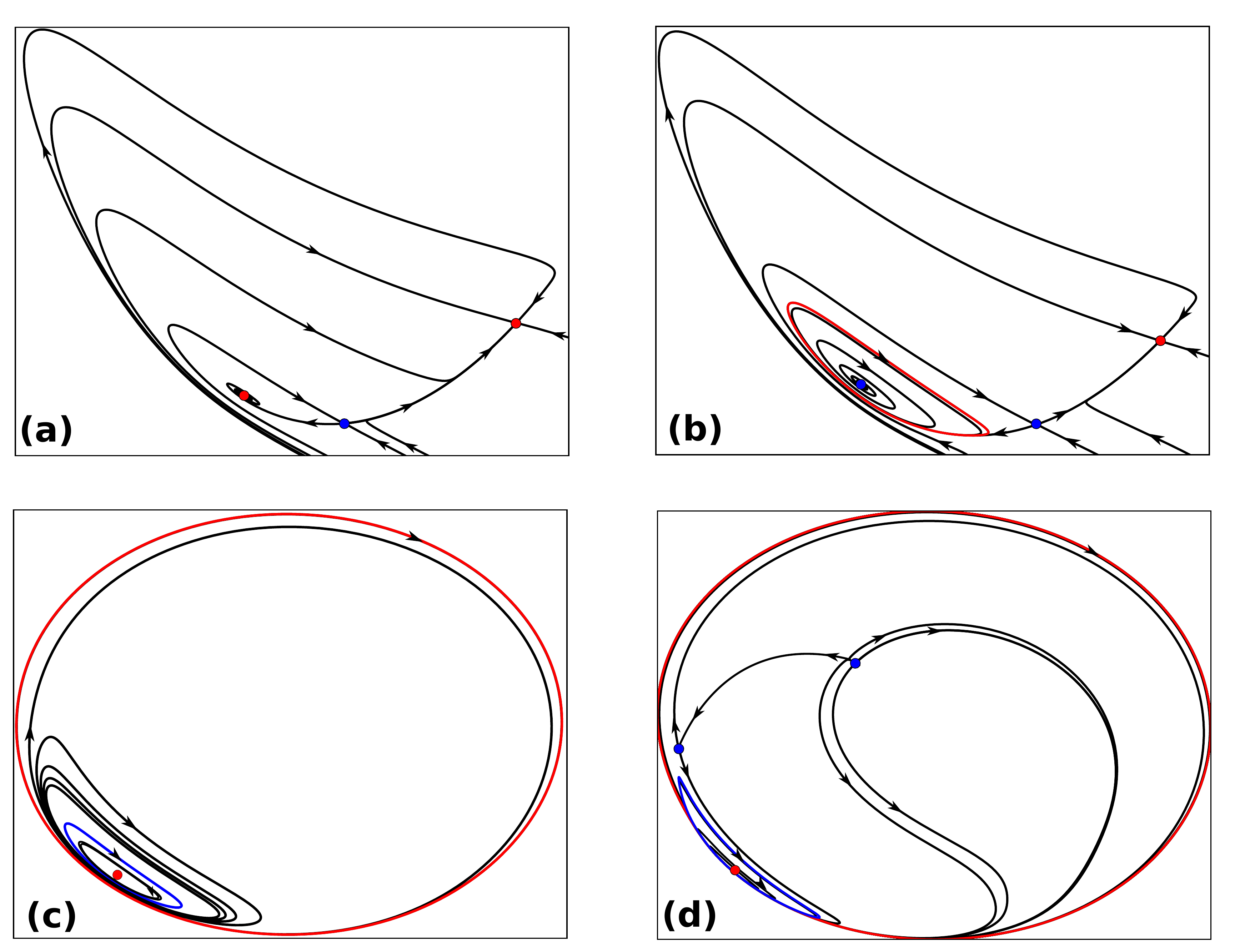}}
\caption{{{(Color online) Phase portraits of the dynamics projected onto the complex plane of the order parameter for bistability regions ($\text{II}_{a}$, $\text{IV}_{a}$, $\text{II}_{b}$, and $\text{IV}_{b}$, shown in (a),(b),(c), and (d), respectively). Initial transients were ignored, and curves were plotted once an effective two-dimensional dynamics was seen. The colors red and blue denote stable and unstable fixed points and cycles, respectively. Panels have been rotated and scaled for clarity.}}} 
\label{fig:PhasePortrait} 
\end{figure}

\begin{figure}[th]
\includegraphics[scale=0.43]{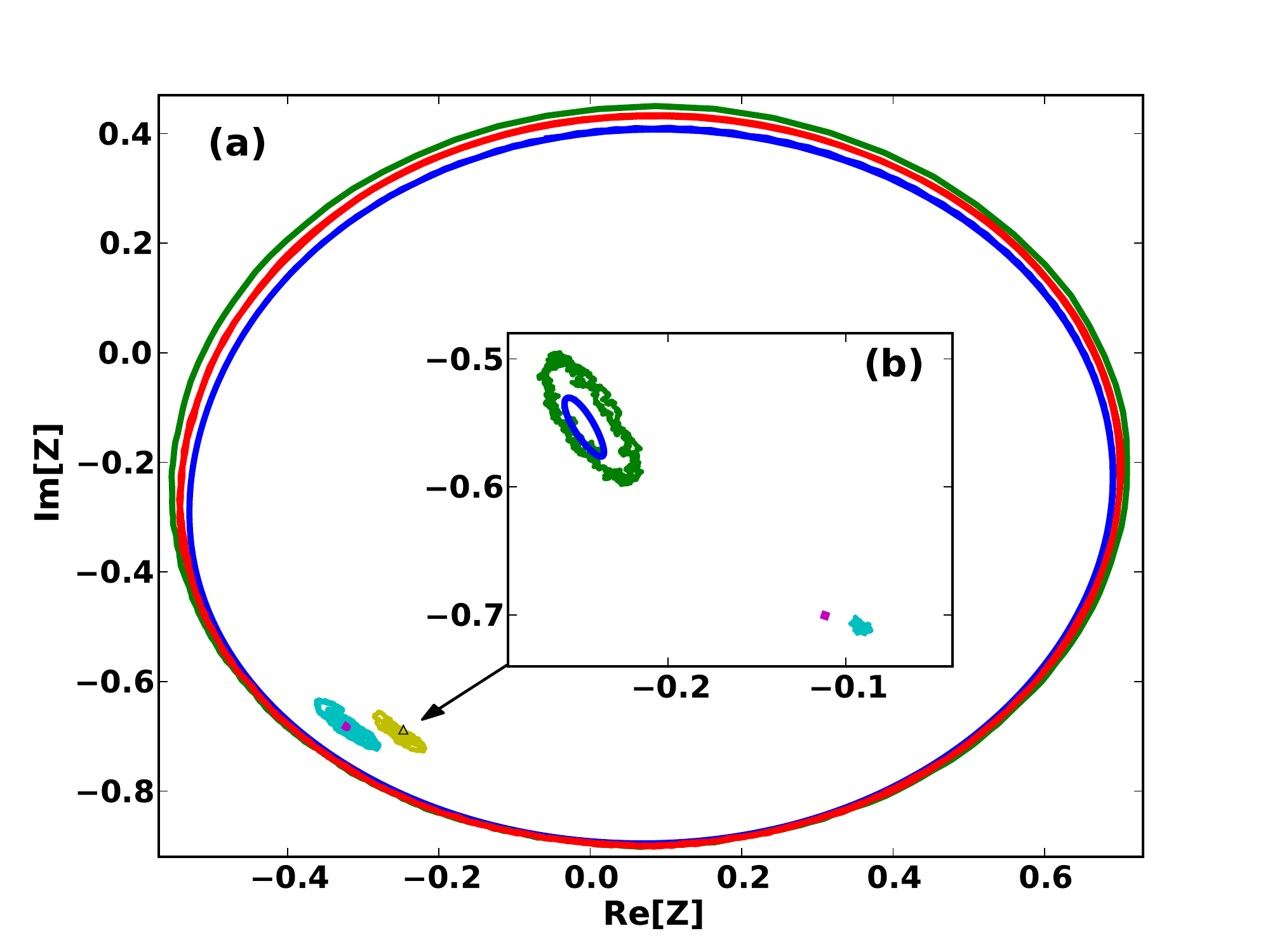}
\caption{{{(Color online) Comparison of bistability of mutual and driven synchronization on heterogeneous and homogeneous networks. (a) Stable states for a network with power-law degree distribution: $\mathcal{E}=7.1$, $\Delta=8.0$, $\mathcal{J}=2.0$, $s=2.0$, and $K_{cut}=200$; the order parameter \eqref{eq:OderParamFiniteN} is shown for the mean-field cycle (blue/black), annealed cycle (green/dark grey), configuration-model cycle (red/medium gray), mean-field equilibrium (magenta/medium gray point), annealed equilibrium (cyan/light gray), and configuration-model equilibrium (yellow with triangle/light gray), with good agreement among the respective states (region $\text{II}_{b}$ in Fig. \ref{fig:Bifurcation} (b)). Networks consist of 30,000 nodes. (b) Analogous plot for a Poisson degree distribution network with the same average degree as (a) and with $\mathcal{J}=0.75$; $\mathcal{E}=2.27$, $\Delta=2.1216$ for the mean-field, and $\mathcal{E}=2.3$, $\Delta=2.134$ for the annealed (region $\text{IV}_{a}$ in Fig. \ref{fig:Bifurcation} (a))$\;$\cite{F3}. The arrow indicates where (b) can be found in $z$'s complex plane for comparison with (a).}}} 
\label{fig:OP} 
\end{figure}

Another important difference between heterogeneous and homogeneous behavior concerns the bistability of driven and mutually synchronized states. Phase portraits are given in Fig.\ref{fig:PhasePortrait}, projected onto the order parameter, which demonstrate the behavior in important parameter regions. For homogeneous networks, bistability exists in a small region of parameter space, confined between the $C$ and $HC$ bifurcations (i.e., regions $\text{II}_{a}$ and $\text{IV}_{a}$ in Fig.\ref{fig:Bifurcation}(a)). In this case, there is bistability between two states of driven synchronization (region $\text{II}_{a}$ and Fig.\ref{fig:PhasePortrait}(a)), until the $H_{sup}$ is crossed (e.g., crossing $\text{II}_{a}-\text{IV}_{a}$), and bistability between a state of quasi-periodic mutual synchronization and driven synchronization (Fig.\ref{fig:PhasePortrait}(b)) \cite{Childs}. In both cases, the manifolds of the saddle act as a separatrix between the two stable states. In contrast, for heterogeneous networks there is only bistability between a large-amplitude state of mutual synchronization and a single state of driven synchronization. The mutually synchronized state encloses all three fixed points in region $\text{IV}_{b}$ (Fig.\ref{fig:Bifurcation}(b) and Fig.\ref{fig:PhasePortrait}(d)), and exists in an additional region that does not contain a saddle ( $\text{II}_{b}$ in Fig.\ref{fig:Bifurcation}(b) and Fig.\ref{fig:PhasePortrait}(c)). An example comparison of the bistability in finite network simulations for the two types of behavior is shown in Fig.\ref{fig:OP}.         


\subsection{\label{sec:ClusterBehavior} Cluster behavior} 
\begin{figure}[!th]
\raggedleft{\includegraphics[scale=0.475]{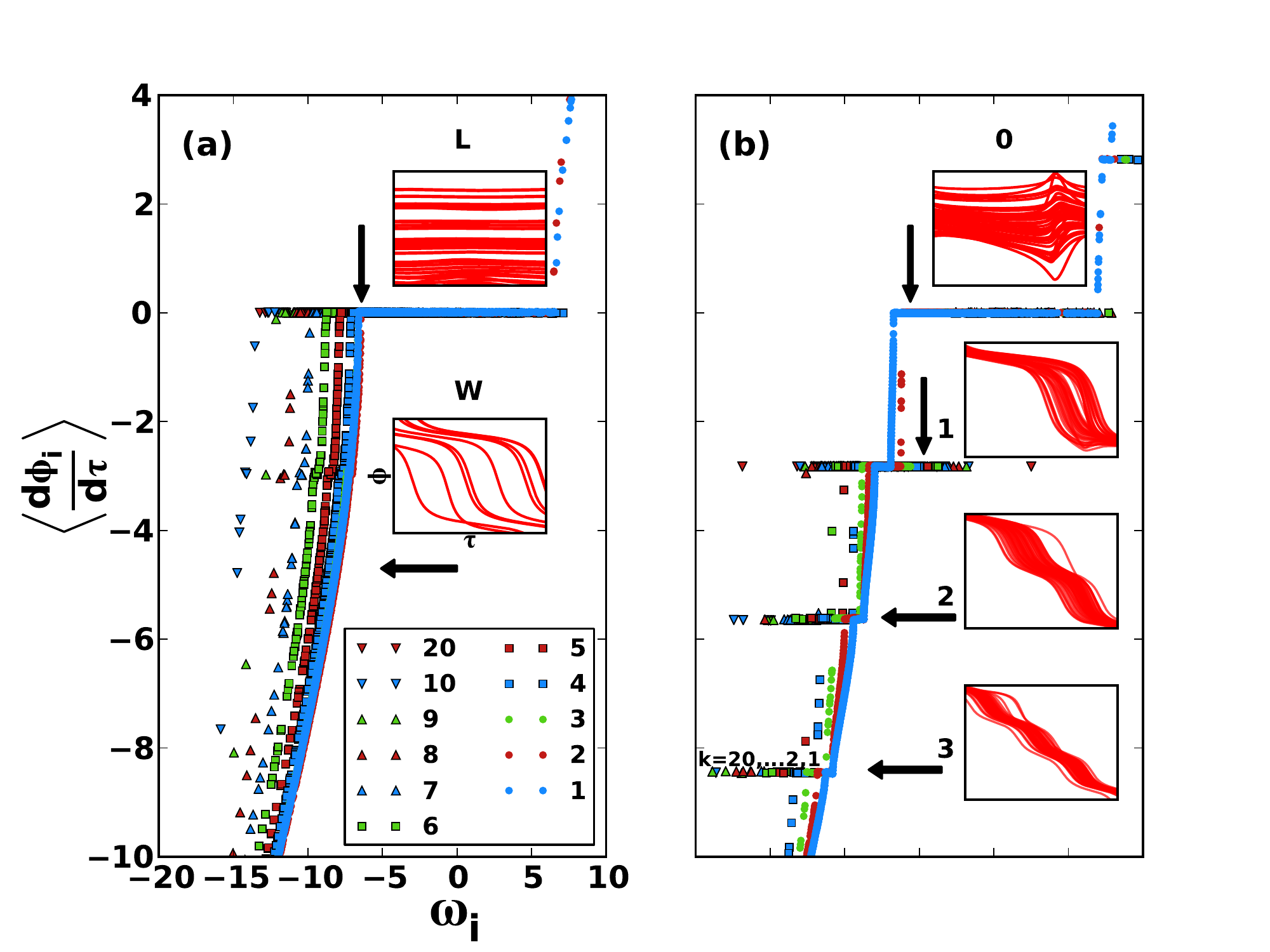}}
\caption{{{(Color online) The average velocities for a network with a power-law degree distribution shown versus natural frequency. (a) A stable state of driven synchronization. (b) A stable state of mutual synchronization. The parameters are $\mathcal{E}=7.1$, $\Delta=8.0$, $\mathcal{J}=2.0$, $s=2.0$, $K_{cut}=200$, and $N=30000$ at which (a) or (b) can be realized, given appropriate initial conditions (region $II_{b}$ in Fig.\ref{fig:Bifurcation}(b)). The inset panels for (b) show $\phi_{i}$ vs. $\tau$ over one cycle with ranges [0,-$\pi$],  [0,-2$\pi$],  [0,-4$\pi$], and [0,-6$\pi$]  for the plateau numbers $n=0,1,2$, and $3$, respectively. The node types for (a) are labeled ``L" for locked ``W'' for winding, and appear next to the inset panels. Also, arrows indicate which cluster of oscillators are shown. A color legend for degree is given in (a).}}} 
\label{fig:Velocity} 
\end{figure}

Finally, we are interested in how the states and transitions discussed in the previous sections appear at a finer scale of resolution: the dynamics of oscillator clusters in the network. For stable states of driven synchronization, we find a single macroscopic cluster of phase-locked nodes, which are entrained to the driving and are stationary in the co-moving frame (labeled ``L" in Fig.\ref{fig:Velocity}(a)). This cluster is comprised of oscillators that have natural frequencies near the driving, with frequency ranges for degree classes that typically increase with degree, so that higher degree classes are able to stabilize a broader range of frequencies. Moreover, nodes with natural frequencies outside of their degree class's locked range have average velocities (time average of Eq.\eqref{eq:ForcingCMFram}) that are monotonically increasing with the displacement from that range, and thus lap the driving field continually with disperse phases from one another. We therefore call these oscillators ``winding" (labeled ``W" in Fig.\ref{fig:Velocity}(a)). The average velocities of phase-locked and winding nodes are shown in Fig.\ref{fig:Velocity}(a) for a driven state as functions of their degrees and natural frequencies.  

On the other hand, a stable state of mutual synchronization has a macroscopic cluster of nodes which lap the field together at some emergent average velocity. In addition, there exist other large ``plateau" clusters of higher harmonics with average velocities that are integer multiples of the fundamental velocity, and therefore lap the driving field $2,3,4...$ times in one network cycle (labeled as $1,2...$ in Fig.\ref{fig:Velocity}(b)). Collectively these harmonic plateaus drive phase-trapped nodes at a frequency equal to the fundamental velocity, causing them to wobble around the driving-field, but with average velocity zero (labeled $0$ in Fig.\ref{fig:Velocity}(b)). The last group of oscillators, which are between the plateaus, wind with average velocities that grow monotonically with the displacement from a given plateau, and have disperse phases. This picture is consistent with general results for Kuramoto models, in which devil's staircases do not appear, and velocities strictly increase between plateaus \cite{Bak, Engelbrecht}. Fig.\ref{fig:Velocity}(b) shows a typical velocity profile for a mutually synchronized state. 
\begin{figure}[th]
\includegraphics[scale=0.32]{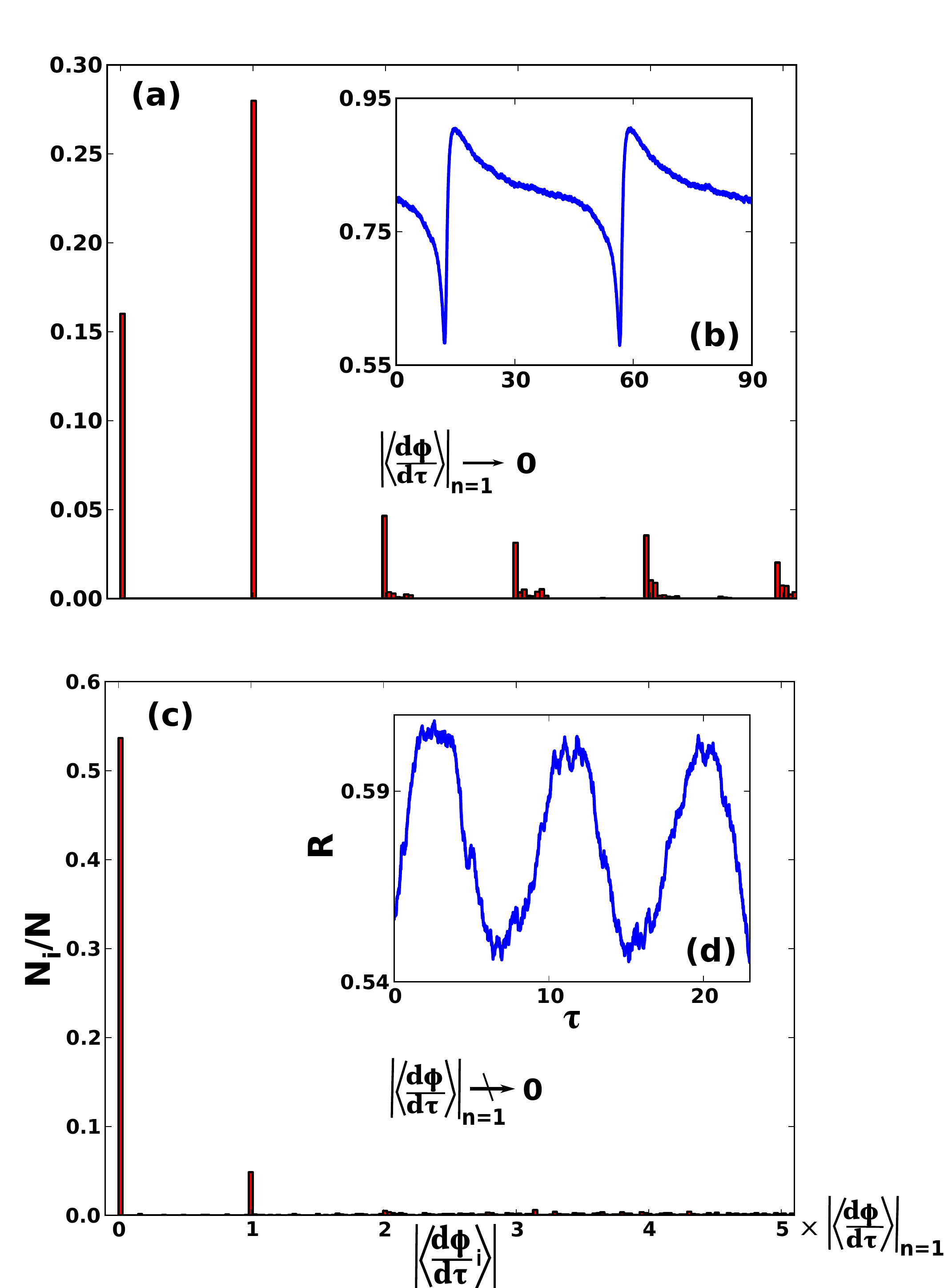}
\caption{{{(Color online) (a) A histogram of the average speeds for a network with a power-law degree distribution just below a $SNIPER$ transition. The parameters are $\mathcal{E}=3.61$, $\Delta=4.0$, $\mathcal{J}=2.0$, $s=2.0$, $K_{cut}=200$, and $N=30000$. We can see that roughly 30\% of the network becomes de-pinned and forms the first plateau, which has speed ($|\left<\frac{d\phi}{d\tau}\right>|_{n=1} \rightarrow0$). (b) The magnitude of the order parameter as a function of time, displaying relaxation dynamics as the network slows down in the neighborhood of a driven state that vanished in the transition. (c) Analogous histogram for a Poisson degree distribution network with the same average degree as (a), but just below a $H_{sup}$ transition with parameters $\mathcal{E}=2.41$, $\Delta=2.6$, $\mathcal{J}=0.75$. In this case, only 5\% of the network occupies the first plateau, which has non-zero speed as the transition is approached. (d) Analogous plot to (b), showing the small amplitude, fast dynamics produced by crossing the $H_{sup}$.}}} 
\label{fig:SNIPER} 
\end{figure} 

Since we find that an important  difference between driven and mutually synchronized states is the appearance of plateaus in the velocity profile, we would like know how the plateaus are occupied when crossing the key transitions\cite{Sakaguchi}. For instance, in crossing over a $SNIPER$ transition, we find that the plateaus of the mutually synchronized state emerge from the phase-locked cluster of a driven state, as a finite fraction of locked nodes with natural frequencies near the average break away from the external field (shown in Fig.\ref{fig:SNIPER}(a-b)). This is consistent with the de-pinning, continuous frequency and discontinuous amplitude appearance predicted by the mean-field dynamics. Conversely when crossing over the $H_{sup}$, we find that a small stable cluster of winding nodes in a driven state, with natural frequencies near the average, coalesce around the same average velocity $\eqref{eq:Hopf}$, and form the first plateau. As the transition is approached the size of each plateau goes to zero (shown in Fig.\ref{fig:SNIPER}(c-d)). This produces the discontinuous frequency and continuous amplitude limit cycle with quasi-periodicity described by the mean-field $H$ bifurcation. Different still, when crossing over the $LPC$, we find that a large group of nodes, which could form the winding and locked clusters of a driven state, can coalesce around an average velocity instead (given appropriate initial conditions). This produces an additional stable state of mutual synchronization that is bistable with the driven state, and has plateaus that are disproportionately occupied by high degree classes. The velocities and order parameter dynamics are compared in Fig.\ref{fig:Velocity} and Fig.\ref{fig:OP}(a) for these bistable states, respectively. 
 \section{\label{sec:Conclusion} CONCLUSION}
In this work we have studied the periodically driven Kuramoto model on random networks with a given degree distribution. A low-dimensional description was found, and a stability and partial bifurcation analysis developed, which allowed us to predict many of the states and transitions of the model for sufficiently weak coupling between nodes\cite{F2}. In particular we found a Takens-Bogdanov-Cusp (TBC) singularity, appearing for power-law degree distribution networks as the degree exponent was lowered, which separated branches of heterogeneous and homogeneous network behavior. The unfolding of this singularity was used to uncover important dynamical transitions including:  Saddle-Node-Infinite-Period, Hopf, and Limit-Point-of-Cycles ($LPC$), as well as multiple bistability regions that differed for the network types. Interestingly, we found that heterogeneous networks do not support bistability of driven synchronized states or bistability of quasi-periodic  synchronized states and driven states (which is the case for homogeneous networks), but only bistability of large amplitude mutually synchronized and driven states. Moreover, we discovered that the $LPC$ transition for the heterogeneous branch occurs with phase-slip dynamics for nodes with high degree and phase-trapped dynamics for nodes with low degree (on average), implying a new route to mutual synchronization for driven heterogeneous networks which allows for qualitatively different behavior depending on a node's degree. In addition, the structure of synchronization clusters for mutual and driven states was discussed and their transitions associated with bifurcations. 

Still, we have yet to resolve all of the transitions associated with unstable cycles in the heterogeneous case (which could inform other interesting features of the dynamics), and the full unfolding of network bifurcations in the strong coupling region. Moreover, many real networks of interest have richer architecture than the simple degree heterogeneity discussed here: such as modular, fractal, and multi-scale structure \cite{Song,Hindes}. The effects of these features on network synchronization are interesting subjects for future work. Finally, the control of complex networks is of immense interest, both theoretical and practical. Our results can offer insight into the problem of controlling disordered oscillator networks.       
   
\section*{\label{sec:Ack}ACKNOWLEDGMENTS}
This work was supported by the Science and Technology Directorate of the U.S.  Department of Homeland Security via the interagency agreement no.\ HSHQDC-10-X-00138. We thank David J. Schneider and John Guckenheimer for useful discussions. 


\begin{thebibliography}{9}
\bibitem{F1} This will be satisfied for virtually any $\mathcal{J} >0$, for random networks with power law degree exponents less than $3$ and large degree cut-offs.
\bibitem{F2} All of the unstable cycle bifurcations in the heterogeneous case have not yet been resolved. Furthermore for power law networks with $k=1,2...K_{cut}$, when $\mathcal{J} \gtrsim2.5$, additional $SN$ curves emerge which complicate the unfolding shown in Fig.\ref{fig:Bifurcation} (See Sec. \ref{sec:BifurcationAnalysis}).
\bibitem{F3} The annealed approximation for configuration model networks is less quantitatively accurate for networks with small average neighbor-degrees (such as the Poisson distributed network shown in Fig.\ref{fig:OP} (b)). This has made it difficult to verify bistability unambiguously for these networks, because the region of parameter space where it is predicted to exist is small (Sec.\ref{sec:KeyTransitions}).
\end{thebibliography}
\end{document}